\documentclass[aps,prl,superscriptaddress,showpacs,twocolumn]{revtex4-1}
\usepackage[left=1.7cm,right=1.7cm,bottom=1.7cm,top=1.7cm,ignoreall]{geometry}
\usepackage[utf8]{inputenc}
\usepackage{graphicx}
\usepackage{amsmath}
\usepackage{amssymb}
\usepackage{xspace}
\usepackage{bm}
\usepackage{color}
\usepackage[hyperindex=true]{hyperref}
\hypersetup{linktocpage,colorlinks=true,citecolor=blue,linkcolor=blue}
\usepackage{empheq}
\usepackage{braket}

\newcommand*{\Scale}[2][4]{\scalebox{#1}{$#2$}}%

\newcommand{\blue}[1]{{\color{black} #1}}

\begin{document}

\title{Unstable and stable regimes of polariton condensation}

\author{F. Baboux}
\thanks{florent.baboux@c2n.upsaclay.fr}
\affiliation{Centre de Nanosciences et de Nanotechnologies, CNRS, Univ. Paris-Sud, Université Paris-Saclay, C2N–Marcoussis, F-91460 Marcoussis, France}
\affiliation{Laboratoire Matériaux et Phéenomènes Quantiques, Université Paris Diderot, CNRS-UMR 7162, Paris 75013, France}

\author{D. De Bernardis}
\affiliation{TQC, Universiteit Antwerpen, Universiteitsplein 1, B-2610 Antwerpen, Belgium}
\affiliation{INO-CNR BEC Center and Dipartimento di Fisica, Università di Trento, I-38123 Povo, Italy}
\affiliation{Vienna Center for Quantum Science and Technology, Atominstitut, TU Wien, 1040 Vienna, Austria}

\author{V. Goblot}
\affiliation{Centre de Nanosciences et de Nanotechnologies, CNRS, Univ. Paris-Sud, Université Paris-Saclay, C2N–Marcoussis, F-91460 Marcoussis, France}

\author{V.N. Gladilin}
\affiliation{TQC, Universiteit Antwerpen, Universiteitsplein 1, B-2610 Antwerpen, Belgium}

\author{C. Gomez}
\affiliation{Centre de Nanosciences et de Nanotechnologies, CNRS, Univ. Paris-Sud, Université Paris-Saclay, C2N–Marcoussis, F-91460 Marcoussis, France}

\author{E.~Galopin}
\affiliation{Centre de Nanosciences et de Nanotechnologies, CNRS, Univ. Paris-Sud, Université Paris-Saclay, C2N–Marcoussis, F-91460 Marcoussis, France}

\author{L.~Le~Gratiet}
\affiliation{Centre de Nanosciences et de Nanotechnologies, CNRS, Univ. Paris-Sud, Université Paris-Saclay, C2N–Marcoussis, F-91460 Marcoussis, France}

\author{A.~Lema\^itre}
\affiliation{Centre de Nanosciences et de Nanotechnologies, CNRS, Univ. Paris-Sud, Université Paris-Saclay, C2N–Marcoussis, F-91460 Marcoussis, France}

\author{I. Sagnes}
\affiliation{Centre de Nanosciences et de Nanotechnologies, CNRS, Univ. Paris-Sud, Université Paris-Saclay, C2N–Marcoussis, F-91460 Marcoussis, France}

\author{I. Carusotto}
\affiliation{INO-CNR BEC Center and Dipartimento di Fisica, Università di Trento, I-38123 Povo, Italy}

\author{M. Wouters}
\affiliation{TQC, Universiteit Antwerpen, Universiteitsplein 1, B-2610 Antwerpen, Belgium}

\author{A. Amo}
\affiliation{Centre de Nanosciences et de Nanotechnologies, CNRS, Univ. Paris-Sud, Université Paris-Saclay, C2N–Marcoussis, F-91460 Marcoussis, France}

\author{J. Bloch}
\affiliation{Centre de Nanosciences et de Nanotechnologies, CNRS, Univ. Paris-Sud, Université Paris-Saclay, C2N–Marcoussis, F-91460 Marcoussis, France}

\begin{abstract}

\blue{Modulational instabilities play a key role in a wide range of nonlinear optical phenomena, leading e.g. to the formation of spatial and temporal solitons, rogue waves and chaotic dynamics.
Here we experimentally demonstrate the existence of a modulational instability in condensates of cavity polaritons, arising from the strong coupling of cavity photons with quantum well excitons. For this purpose} we investigate the spatiotemporal coherence properties of polariton condensates in GaAs-based microcavities under continuous-wave pumping.
The chaotic behavior of the instability results in a strongly reduced spatial and temporal coherence and a significantly inhomogeneous density.
Additionally we show how the instability can be tamed by introducing a periodic potential so that condensation occurs into negative mass states, leading to largely improved coherence and homogeneity. 
\blue{These results pave the way to the exploration of long-range order in dissipative quantum fluids of light within a controlled platform.}

\end{abstract}

\maketitle

\section{Introduction}

\blue{Modulational instabilities are a widesperead feature of nonlinear wave systems, whereby small perturbations get amplified and grow exponentially with time. They manifest in numerous branches of physics ranging from hydrodynamics \cite{Benjamin67} to nonlinear optics \cite{Tai86}, plasma physics \cite{Amiranoff92} and cold atom gases \cite{Strecker02}.
In optical systems such as fibers \cite{Tai86,Wright16}, waveguide arrays \cite{Stegeman99}, lasers \cite{Hegarty99,Sciamanna15,Selmi16} and amplifiers \cite{Hammani08}, modulational instabilities have been shown to deeply affect the temporal or spatial properties of light fields, leading e.g. to the break-up of uniform beams into trains of solitons \cite{Mollenauer80,Stegeman99}, the generation of filaments \cite{Carvalho96} or the emergence of extreme intensity fluctuations known as optical rogue waves \cite{Solli07,Selmi16}.
Chaotic dynamics related to instabilities have been especially studied in lasers \cite{Arecchi82,Selmi16} with envisioned applications in random number generation and optical sensing \cite{Sciamanna15}.

Cold atom condensates, where the sign and strength of nonlinearities can be dynamically tuned using Feshbach resonances \cite{Roberts01,Donley01}, have also been a fruitful playground for the experimental investigation of instabilities.
An homogeneous cold atom condensate becomes unstable in presence of attractive interactions, which make the condensate collapse so as to minimize its potential energy~\cite{StringariBook,Stoof94}. In presence of a confining potential however, zero-point kinetic energy competes with interactions and a condensate can persist up to a given critical density \cite{Gerton00} after which the condensate implodes \cite{Donley01,Lahaye08} or breaks into a train of solitons \cite{Strecker02,Nguyen17}. Inspired by dispersion management techniques used in fiber optics to generate solitons or gap solitons \cite{BoydBook,ButcherBook,DeSterke94}, self-bound condensate droplets have also been observed for repulsively interacting atoms when the atomic effective mass is turned to negative by means of an optical lattice \cite{Anker05} or of a spin-orbit coupling \cite{Khamehchi17}. For non-equilibrium systems however -- such as in presence of drive and dissipation -- the fate of the instability of bosonic condensates has been little explored experimentally so far.}

Polaritons, arising from the strong coupling of quantum well excitons and cavity photons \cite{Carusotto13}, are an appealing candidate to address this question. These quasi-particles indeed combine non-equilibrium properties with substantial interactions. Direct interactions between polaritons are of repulsive nature \cite{Ciuti98}, which should preclude the aforementioned instability scenario. However, under the widely used non-resonant pumping scheme, interactions between polaritons and the reservoir cloud of uncondensed excitons need also to be considered. This polariton-reservoir coupling can be shown to give effective \textit{attractive} interactions between condensed polaritons~\cite{Smirnov14,Liew15}. When these effective interactions overcome the direct repulsive interactions between polaritons, the condensate is expected to enter a modulationally unstable regime \cite{Wouters07} characterized by a turbulent steady state \cite{Cuevas11,Smirnov14,Bobrovska14,Bobrovska15,Liew15}. This behavior strongly contrasts the collapse scenario of attractively interacting cold atom gases and is another intriguing example of the rich non-equilibrium physics of driven-dissipative polariton condensates.

While predicted a decade ago \cite{Wouters07} this instability of polariton condensates has been hindered in previous experimental reports, except for a recent study using organic cavities under pulsed excitation \cite{Bobrovska16}. 
Here, we evidence the presence of instabilities in the most typical case of \textit{inorganic} cavities and under \textit{continuous-wave} (CW) pumping, and we highlight how their signatures are markedly distinct from the ones observed in Ref. \cite{Bobrovska16}. We also demonstrate a method to suppress the polariton instability.
For this purpose, we investigate the coherence properties of polariton condensates in 1D and 2D GaAs-based cavities in which the sign of the polariton effective mass can be changed. 
When condensation occurs in a positive mass state, we observe an unstable steady state regime characterized by a strongly reduced spatial and temporal coherence, and a sizable density inhomogeneity. When the cavities are spatially patterned into lattices so that condensation occurs in negative mass states, the modulational instability is suppressed and all previous signatures of instability disappear: the condensates reach a stable steady-state with high homogeneity and coherence. This method for suppressing instabilities opens avenues for studying the rich phenomenology of driven-dissipative condensates \cite{Sieberer13,Chiocchetta13,He15,Altman15,Dagvadorj15,Gladilin14,Ji15,Squizzato17}, such as Kardar-Parisi-Zhang (KPZ) universal scalings, in a controlled environment.
\blue{The presented results are relevant to a wide range of photonic plateforms sustaining Bose-Einstein-like transitions, such as vertically emitting laser diodes \cite{Lundeberg07}, photon condensates \cite{Klaers10} and plasmonic structures embedded in cavities \cite{Ramezani17}. Moreover, an intense research effort is currently focused on polaritons in two-dimensional electron gases \cite{Sidler17,Ravets18} or transition metal dichalcogenides \cite{Liu15,Dufferwiel15} embedded in cavities. Our proposal of using negative mass states to facilitate the establishment of long-range order could prove useful to achieve stable condensation in these novel platforms.}

\section{Theoretical framework}

We start with a theoretical description of polariton condensation showing the physical origin of the reservoir-induced modulational instability. Ignoring the spin degree of freedom, the condensate wavefunction $\psi(\textbf{r},t)$ can be described by a generalized Gross-Pitaevskii equation coupled to a rate equation for the exciton reservoir density $n_R(\textbf{r},t)$ \cite{Wouters07}:
\begin{empheq}[left={\empheqlbrace}]{alignat=2}
& \Scale[0.92]{
             i\hbar \frac{\partial \psi }{\partial t}=\left[ -\frac{\hbar ^{2}\Delta }{2m}%
             +g\left\vert \psi \right\vert ^{2}+2g_{R}n_{R}+\frac{i\hbar }{2}\left(
             Rn_{R}-\gamma \right) \right] \psi}  \\
& \Scale[0.92]{
             \frac{\partial n_{R}}{\partial t}=P(\textbf{r})-(\gamma _{R}+R\left\vert \psi
             \right\vert ^{2})n_{R}%
             }
\end{empheq}
where $P(\textbf{r})$ is the pumping rate, $m$ is the effective polariton mass at the condensate energy, $\gamma$ and $\gamma_R$ are the polariton and exciton loss rates and $R$ is the relaxation rate of the reservoir into the condensate. $g$ and $g_R$ are positive and describe the repulsive polariton-polariton and polariton-reservoir interaction constants.

The widely used adiabatic approximation assumes that the reservoir follows instantaneously the condensate dynamics and is expected to be accurate if the reservoir decay time $\gamma_R^{-1}$ is the fastest time scale \cite{Remark}. Under this condition, Eq. (2) reduces to $n_{R}(\textbf{r})=P(\textbf{r})/( \gamma _{R}+R\left\vert \psi \right\vert ^{2})$. Reinjecting into Eq. (1) yields a modified Ginzburg-Landau equation \cite{Carusotto13} in which $g$ is replaced by an effective interaction constant \cite{Smirnov14,Liew15}:
\begin{equation}
g_{\rm eff}=g-2 g_{R}\frac{\gamma }{\gamma_R}\frac{P_{\rm th}}{P}
\label{g_eff}
\end{equation}
This means that, as far as the condensate dynamics is concerned, the coupling to the reservoir reduces to an effective interaction between polaritons. The condensate is then dynamically stable for repulsive interactions ($g_{\rm eff}>0$, defocusing effective nonlinearity), and dynamically unstable for attractive interactions ($g_{\rm eff}<0$, self-focusing effective nonlinearity).

This polariton instability can be interpreted as a reservoir-induced modulational instability as follows. A local increase of polariton density (due to a quantum or thermal fluctuation of the condensate, or to pump noise) induces a local depletion of the reservoir density via a spatial hole burning \cite{Wouters07,Estrecho17}. Such a depletion creates a potential well which further attracts the condensate polaritons, making the initial fluctuation to exponentially grow in time. This positive feedback loop is eventually broken by gain saturation and by polariton propagation, so that the density fluctuation is ejected from its initial position and starts moving through the condensate. A turbulent behavior results from the chaotic evolution of several of such density fluctuations \cite{Bobrovska14,Bobrovska15,Liew15}.

In typical polariton experiments \cite{Kasprzak06,Deng07,Manni12,Roumpos12,Nitsche14,Fischer14}, $g \sim 0.05-0.5 \, g_R$ and $\gamma \sim 10-100 \, \gamma_R$, so that the stability condition $g_{\rm eff}>0$ requires high pump powers $P \sim 40-4000 \, P_{\rm th}$ difficult to achieve in practice, in particular for large pump spots in a CW regime. This stability condition can be relaxed by using a relatively small pumping spot ($\sim 2-10 \, \mu$m) \cite{Bobrovska14,Daskalakis15,Bobrovska16}, as was done in most previous experimental studies \cite{Kasprzak06,Deng07,Manni12,Roumpos12,Ohadi12,Nitsche14,Fischer14}.
In that case, the confinement-induced kinetic energy and, possibly, the outwards flow due to the overall density profile competes with attractive interactions so that stable condensates can be achieved even for negative $g_{\rm eff}$. But this method imposes small condensate sizes, which
limits the capacity of polaritons to serve as a platform
for simulating novel driven-dissipative phenomena \cite{Sieberer13,Chiocchetta13,He15,Altman15,Dagvadorj15,Gladilin14,Ji15,Squizzato17}.
A promising alternative are 
the large condensates recently demonstrated in cavities with long 
lifetimes and reduced wedge \cite{Ballarini17}. However, the fact that the condensate is fed by ballistically expanding fast polaritons instead of (almost) immobile excitons introduces a more complex spatial dynamics into the reservoir equation (2), which is likely to modify the universal properties of the transition.

In contrast to all these works, here we demonstrate how the dynamical instability of condensates under CW pumping with a large spot ($\approx 90\,\mu$m) can be tamed by introducing a periodic potential. When condensation occurs into negative mass states, stable condensates of arbitrarily large size (only limited by the available total pump power) can be obtained while maintaining a coupling to an excitonic reservoir of the standard form (2), for which rich KPZ driven-dissipative phase transitions are expected \cite{He15,Altman15,Gladilin14,Ji15,Squizzato17}.

\begin{figure*}[t]
\centering
\includegraphics[width=0.8\textwidth]{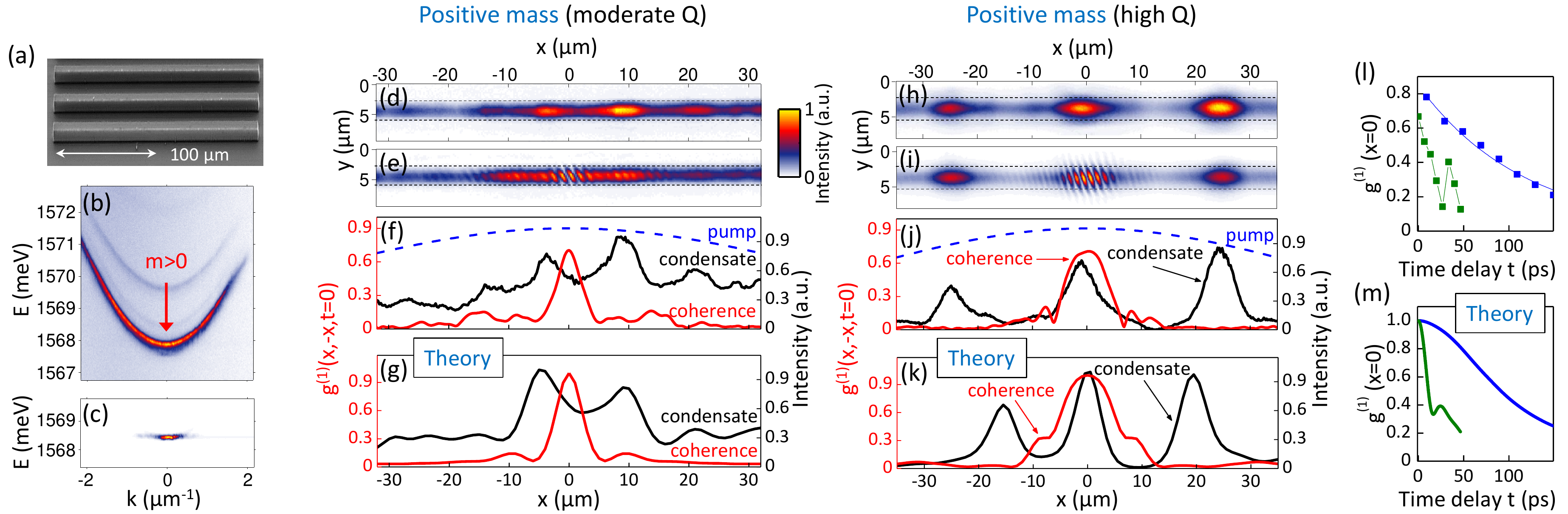}
\caption{
(a) Scanning electron micrograph of 1D wire cavities.
(b)-(c) Far field emission (TM polarization) of a moderate $Q$ wire cavity at pump power (b) below and (c) above the condensation threshold ($P = 2 P_{\rm th}$).
(d) Real space image of the condensate.
(e) Interferogram obtained by superposing two mirror-symmetric images of the condensate in a Michelson interferometer.
(f) Measured and (g) calculated spatial coherence $g^{(1)}$ at $t=0$ (red), compared to the spatial profile of the condensate (black) and of the pump spot (blue).
(h)-(k) Same quantities as (d)-(g) obtained in a high $Q$ wire cavity.
(l) Measured and (m) calculated evolution of the $x=0$ temporal coherence, for the moderate (green line) and high  $Q$ wire cavity (blue line).
}
\label{Fig1}
\end{figure*}

\begin{figure}[t]
\centering
\includegraphics[width=\columnwidth]{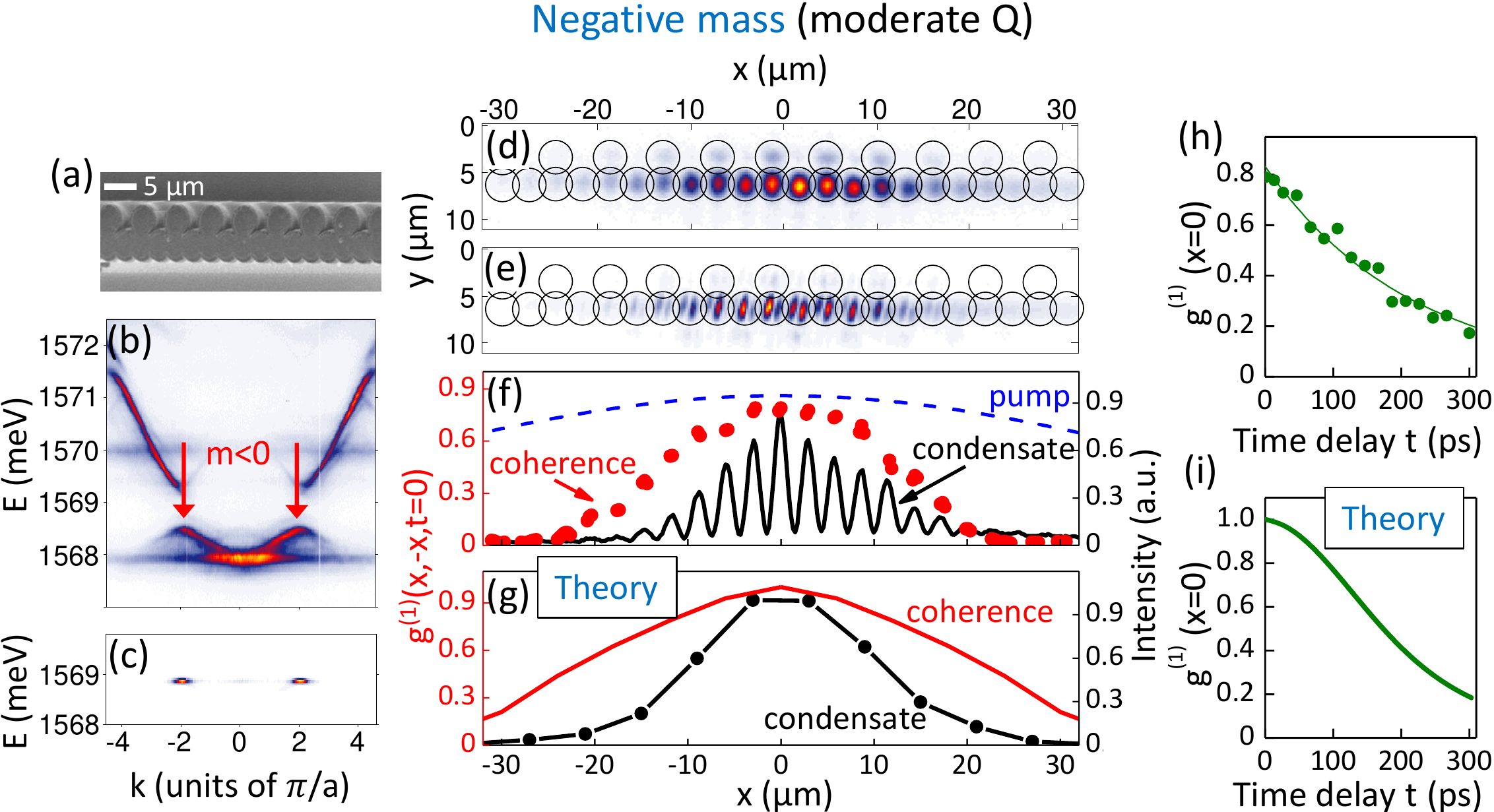}
\caption{
(a) Scanning electron micrograph of a 1D Lieb lattice of micropillars.
(b)-(c) Far field emission (TM polarization) of a moderate $Q$ Lieb lattice at pump power (b) below and (c) above the condensation threshold ($P = 1.5 P_{\rm th}$).
(d) Real space image and (e) interferogram of the condensate.
(f) Measured and (g) calculated spatial coherence $g^{(1)}$ at $t=0$ (red), compared to the spatial profile of the condensate (black) and of the pump spot (blue).
(h) Measured and (j) calculated evolution of the $x=0$ temporal coherence.
}
\label{Fig2}
\end{figure}

\section{Samples and experimental setup}

Our microcavities, grown by molecular beam epitaxy, consist of a $\lambda / 2$ Ga$_{0.05}$Al$_{0.95}$As layer surrounded by two Ga$_{0.8}$Al$_{0.2}$As/Ga$_{0.05}$Al$_{0.95}$As Bragg mirrors. To vary the quality factor we fabricate two sets of cavities: (1) Moderate $Q$ factor with 26 and 30 pairs in the top/bottom mirrors, respectively, yielding a nominal quality factor $Q=30 000$;
(2) High $Q$ factor with 28 and 40 pairs in the top/bottom mirrors, yielding  nominally $Q=70 000$.
For both types of cavities twelve GaAs quantum wells of width 7 nm are inserted in the structure, resulting in a 15 meV Rabi splitting. The planar cavities are studied as such, or patterned into ridges or lattices of coupled micropillars, by e-beam lithography followed with dry etching (down to the GaAs substrate).

In the experiments described below, polaritons are excited non-resonantly with a CW monomode laser tuned to 740 nm. The sample temperature is $6$ K and the cavity-exciton detuning is $-10$ meV (defined as the difference between the lowest-energy cavity mode and the exciton resonance). The polariton emission is collected with a $0.5$ numerical aperture objective and focused on the entrance slit of a spectrometer coupled to a CCD camera. Imaging of the sample surface (resp. the Fourier plane) allows for studying polariton properties in real (resp. reciprocal) space.

\section{Experimental results in 1D cavities}

We first consider 1D cavities, and we investigate polariton condensation
in positive mass states. A wire cavity (width $3$ $\mu$m and length $200$ $\mu$m, see Fig. \ref{Fig1}a) of
moderate $Q$ factor is excited with an elliptical spot of length $90$ $\mu$m (intensity FWHM).
Fig.~\ref{Fig1}b shows the far field emission at very low pump power (for the TM polarization, i.e. parallel to the wire axis), evidencing a parabolic-like dispersion (near $k\!=\!0$) with an effective mass $m\simeq 5 \times 10^{-5} \, m_0$ ($m_0$ is the free electron mass).
When increasing the pump power, stimulated scattering causes the emission to collapse into a narrow spectral line \cite{Kasprzak06} centered at $k=0$, as seen in the spectrum of Fig. \ref{Fig1}c obtained at $P=2 P_{\rm th}$.
The real space image of the resulting polariton condensate (Fig. \ref{Fig1}d) and the corresponding spatial profile (Fig. \ref{Fig1}f, black line) reveal inhomogeneities in the condensate density, with a typical contrast (ratio between maximum and minimum density) $\mathcal{C}\simeq 2$ at the center of the pump spot.

To investigate the coherence properties of the condensate we employ Michelson interferometry \cite{Kasprzak06}. We superpose the condensate image with its mirror symmetric, obtained by reflection on a retroreflector, so that each point $x$ of the condensate interferes with the point located at $-x$. The corresponding interferogram is shown in Fig. \ref{Fig1}e, for zero temporal delay ($t=0$) between the two arms of the interferometer. By extracting the fringe visibility through Fourier analysis, we obtain the first order spatial coherence $g^{(1)}(x,-x, t=0)$, which is plotted in Fig. \ref{Fig1}f (red line). The measured coherence extends over a much shorter length scale than the condensate density, with a coherence length (at $1/e$) of $l_c=6$ $\mu$m. This short spatial coherence is a first hint of the presence of an instability and of the consequent turbulent behavior. To gain further insight into this phenomenon, we investigate the temporal coherence. Figure \ref{Fig1}l (green line) shows the evolution of $g^{(1)}(x=0)$ when scanning the temporal delay $t$ of the interferometer. The decay is non-monotonic (a revival is seen near $t=40$ ps, see Supplementary Material) and the envelope decays within a coherence time $\tau_c \simeq 50$ ps.

The density inhomogeneity (Fig. \ref{Fig1}f) suggests that disorder is playing an important role in the experiment, leading to a significant spatial modulation of the condensate. 
We expect this effect to be amplified if the ratio between the disorder amplitude and the polariton linewidth increases (see discussion at the end of Section 5). To test this dependence
we perform the same set of experiments in a wire of high $Q$ factor but similar disorder strength.
The real space image (Fig. \ref{Fig1}h) and spatial profile (Fig. \ref{Fig1}j, black) of a condensate at $P=2 P_{\rm th}$ indeed reveals a much stronger density inhomogeneity, with a typical contrast $\mathcal{C}\sim 10$: the condensate fragments into distinct lobes.
The interferogram (Fig. \ref{Fig2}i) and extracted spatial $g^{(1)}$ (Fig. \ref{Fig2}j, red) demonstrate that there is no mutual coherence between the different lobes of the condensate. Hence, the coherence length is limited to the lobe size and is of the order of $l_c=8$ $\mu$m, comparable to the one of the moderate $Q$ condensate of Fig. \ref{Fig1}f. The coherence time, on the other hand, is here twice longer with $\tau_c \simeq 120$ ps for each lobe (Fig. \ref{Fig1}l, blue line).

Let us now investigate condensation in negative mass states, as it can be achieved by patterning the cavity into a lattice \cite{Jacqmin14,Klembt17,Whittaker17}. Our intuition is here guided by the adiabatic approximation, that led to the derivation of Eq. (\ref{g_eff}) for the effective interaction $g_{\rm eff}$ between polaritons. In this approximation one can easily show that, all other parameters kept the same, inverting the sign of the mass reverses the effect of interactions \cite{BoydBook,ButcherBook,DeSterke94}, suppressing the feedback loop at the heart of the modulational instability.

To test this prediction, we fabricate a 1D Lieb lattice of coupled micropillars \cite{Baboux16}, as shown in Fig. \ref{Fig2}a. We here present data for a moderate $Q$ cavity but we obtained similar results at high $Q$. Each pillar has a diameter of $3$ $\mu$m and the lattice period is $a=5.8$ $\mu$m. The fundamental (s-symmetry) states of the pillars hybridize to form three bands. For our lattice parameters the two lowest are superimposed within the linewidth, but the upper band is well separated, as seen in the far field emission of Fig. \ref{Fig2}b (TM polarization). This upper band shows a negative curvature at the center of the second Brillouin zones ($k=\pm 2\pi /a$, see vertical red arrows), yielding a negative effective mass $m\simeq -4 \times 10^{-5} \, m_0$ nearly equal to the opposite of the mass of the wire cavities studied above \cite{Remark2}.
Fig. \ref{Fig2}c shows the condensate far field emission at $P=1.5 P_{\rm th}$, which is concentrated at the top of the upper band.
The real space image (Fig. \ref{Fig2}d) and corresponding spatial profile (Fig. \ref{Fig2}f, black) show that the condensate possesses a regular Gaussian-like envelope, the disorder only inducing a minor modulation.
The spatial coherence at $t=0$, extracted from the interferogram of Fig. \ref{Fig2}e, is shown in Fig. \ref{Fig2}f (red circles). As the antisymmetric character of the upper band makes the condensate density to vanish in between neighboring pillars, we restricted our spatial sampling of the $g^{(1)}$ coherence function to the center of the pillars.
We observe that the spatial coherence extends over the whole condensate, yielding a coherence length $l_c \simeq 35$ $\mu$m, about $5$ times higher than for positive mass condensates (Figs. \ref{Fig1}f and \ref{Fig1}j).
The temporal coherence $g^{(1)}(x=0,t)$, shown in Fig. \ref{Fig2}h, reveals a slow and monotonic decay with a long coherence time $\tau_c \simeq 210$~ps, four times higher than the one of the wire cavity of same $Q$ factor (Fig.~\ref{Fig1}l, green). This strongly suggests that the condensate is here dynamically stable.

\begin{figure}[t]
\centering
\includegraphics[width=1\columnwidth]{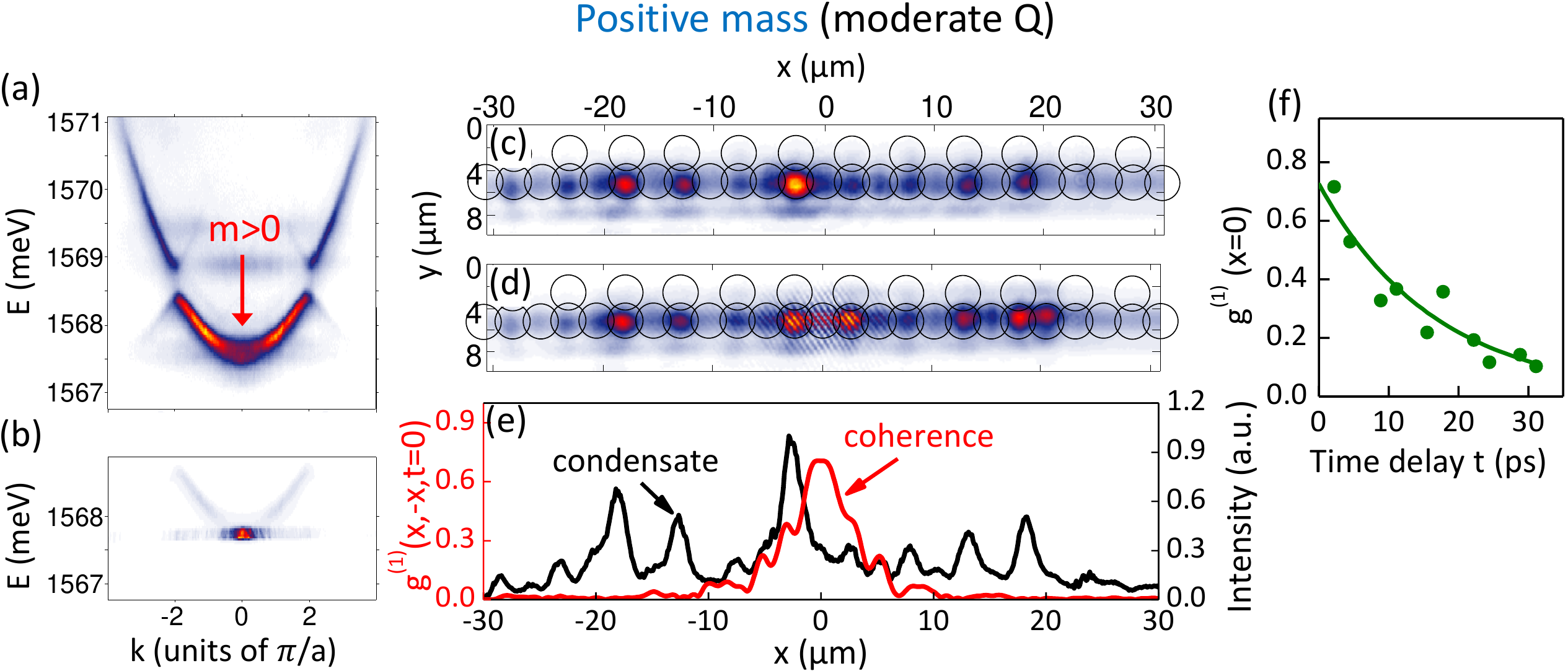}
\caption{
(a)-(b) Far field emission of a moderate $Q$ Lieb lattice at pump power (a) below and (b) above the condensation threshold ($P = 1.5 P_{\rm th}$) when condensation is triggered into a positive mass state.
(c) Real space image and (d) interferogram of the condensate.
(e) Measured spatial coherence $g^{(1)}$ at $t=0$ (red), compared to the spatial profile of the condensate (black).
(h) Measured evolution of the $x=0$ temporal coherence.
}
\label{Fig_Lieb_lower_band}
\end{figure}

To verify that the high coherence and homogeneity are due the negative mass, and not simply to the patterning of the cavity into micropillars, we also consider condensation in a positive mass state of the lattice. For this, we use a lattice with slightly reduced cavity-exciton detuning (-8 meV instead of -10 meV) so as to enhance the relaxation of polaritons towards the fundamental state. In this condition, condensation is triggered at the bottom of the lower energy band, as shown in the far field spectrum of Fig. \ref{Fig_Lieb_lower_band}b (power $P = 1.5 P_{\rm th}$, lattice period $a=5.2$ $\mu$m). This positive mass condensate is highly inhomogeneous (Fig. \ref{Fig_Lieb_lower_band}c and e) and has short spatial coherence (Fig. \ref{Fig_Lieb_lower_band}d and e) \cite{Remark3}, similarly to the positive mass condensates observed in 1D wires (Fig. \ref{Fig1}). Furthermore, the coherence time $\tau_c\simeq 20$ ps (Fig. \ref{Fig_Lieb_lower_band}f) is one order of magnitude smaller than for the negative mass condensates (Fig. \ref{Fig2}h). 
Hence, the patterning of a cavity into a lattice of micropillars is not sufficient to obtain homogeneous and highly coherent condensates. These are only obtained when condensation occurs into negative mass states.

Finally, a further evidence in support of our interpretation is obtained by comparison to the case of condensation in a flat energy band, which we have reported previously \cite{Baboux16}. Since the effective mass is infinite, kinetic energy is suppressed in this case: density fluctuations cannot propagate and the instability mechanism described here is quenched. A condensate with homogeneous profile is thus obtained but strongly reduced spatial coherence reveals the fragmentation of the condensate into elementary plaquettes.

\section{Numerical simulations}

\begin{figure}[t]
\centering
\includegraphics[width=0.9\columnwidth]{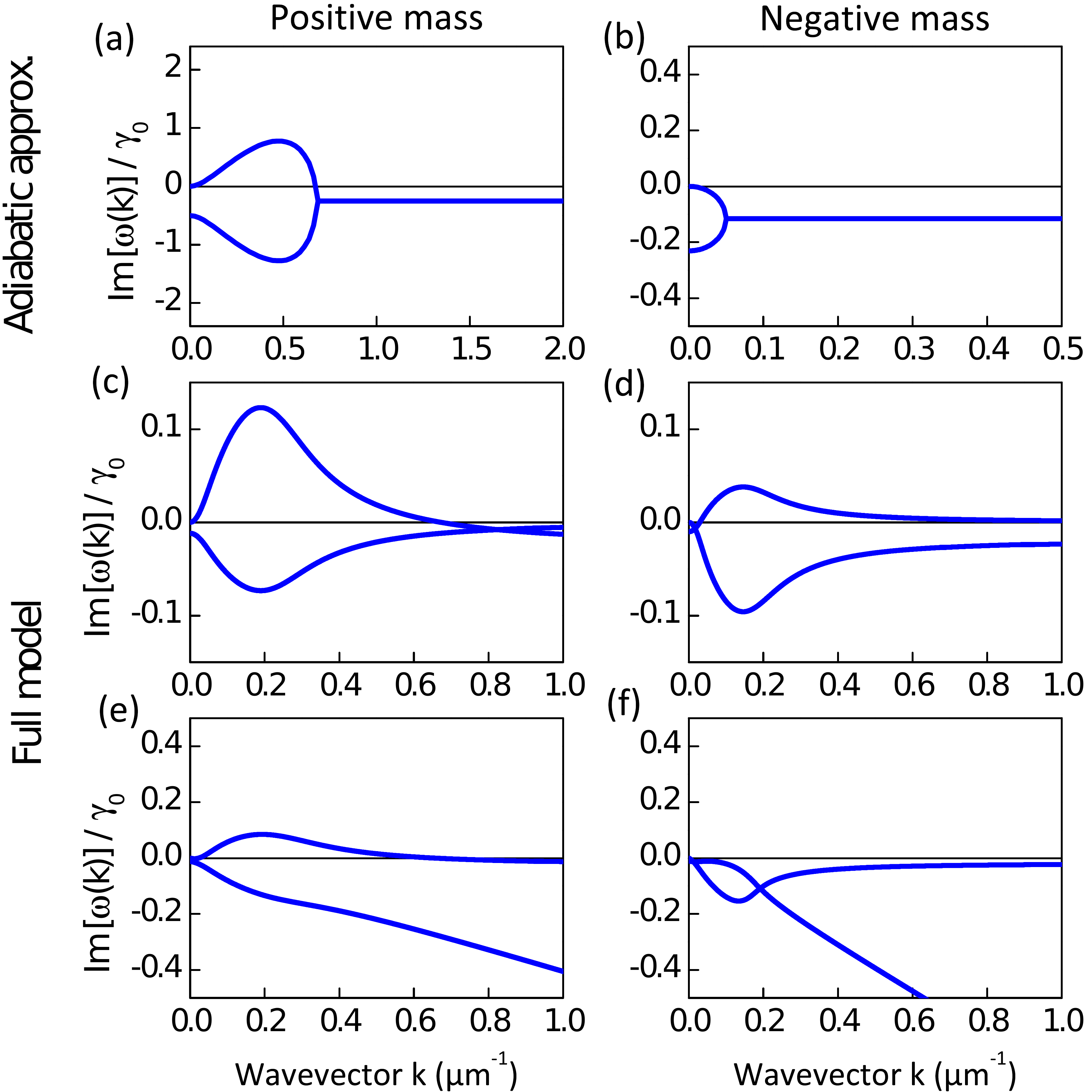}
\caption{
(a)-(b) Imaginary part of the Bogoliubov excitation spectrum calculated under the adiabatic approximation, for a positive (a) and negative (b) mass condensate of moderate $Q$ factor, assuming a constant polariton linewidth $\gamma_0$.
(c)-(f) Same quantities calculated with the full model of coupled Eqs. (1)-(2), with a constant linewidth (c,d) and with a momentum-dependent linewidth (e,f). Formulas and parameters are given in the Supplementary Material.
}
\label{Fig3}
\end{figure}

To get further physical insight into these condensation behaviors, let us now compare these experimental results to theoretical predictions. We start from linear stability analysis of the time-independent steady-state, assuming a spatially homogeneous system. Using parameters taken from the experiment, we calculate the spectrum $\omega(k)$ of the elementary (Bogoliubov) excitations (see Supplementary Material). We first consider the adiabatic approximation, which has been widely used in the literature to describe various polaritonic experiments employing small excitation spots. Figure \ref{Fig3}a shows the imaginary part of the spectrum for the positive mass condensate in the moderate $Q$ wire cavity. 
In the low wavevector region, the upper Goldstone branch takes positive imaginary values: perturbations at these wavevectors are exponentially amplified by the system, corresponding to a modulationally unstable regime of condensation.
Figure \ref{Fig3}b shows the spectrum calculated for the negative mass condensate in the moderate $Q$ lattice: here, all excitation modes have negative imaginary part and are thus exponentially damped, corresponding to a stable regime of condensation.

\begin{figure*}[t]
\centering
\includegraphics[width=0.7\textwidth]{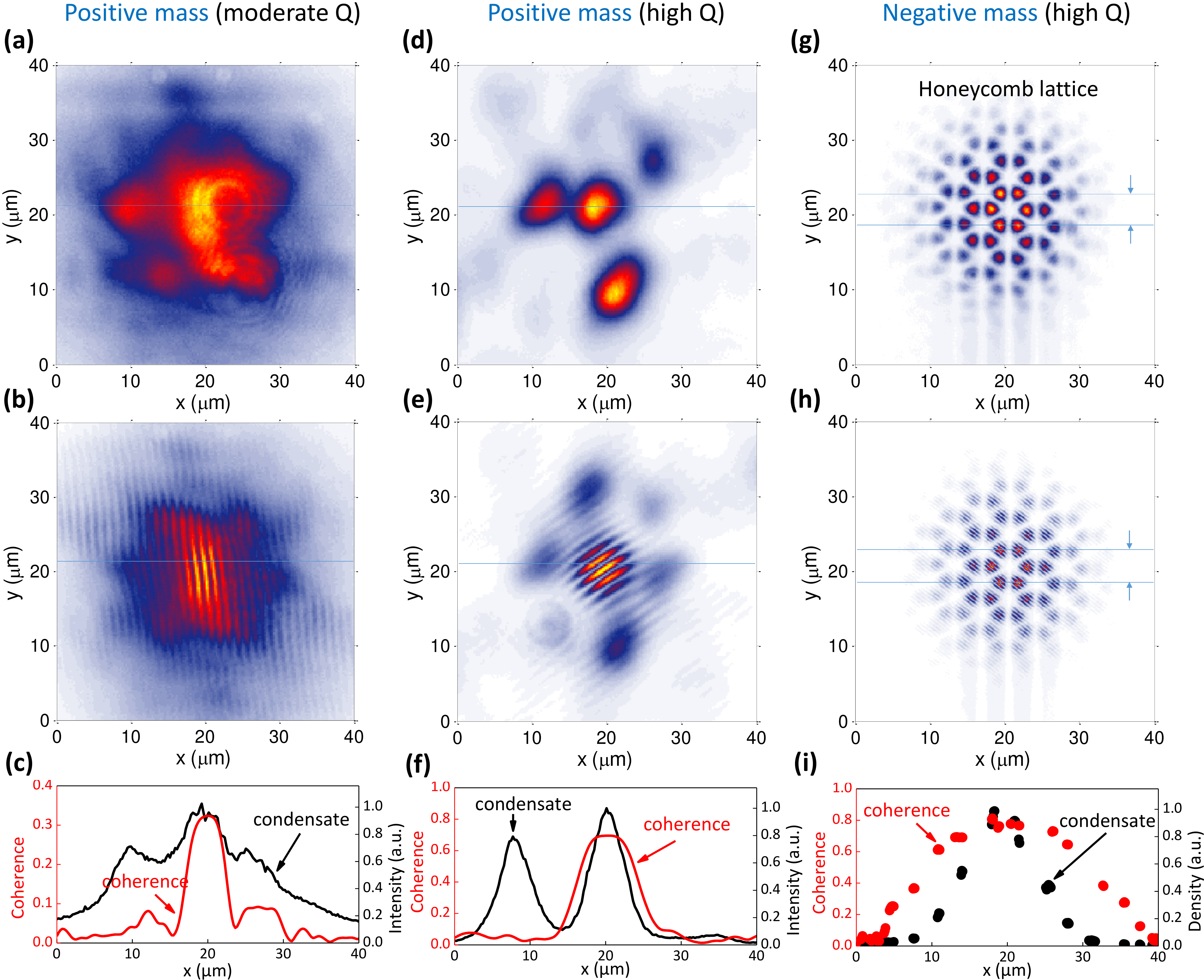}
\caption{Condensation in 2D cavities.
(a) Real space image and (b) interferogram of the condensate in a planar cavity with moderate $Q$ factor.
(c) Cuts of the the condensate density and of the spatial coherence profiles along the horizontal axis indicated in (a)-(b).
(d)-(f) Similar measurements performed on a high $Q$ planar cavity.
(g)-(i) Similar measurements performed on a high $Q$ honeycomb lattice of micropillars.
}
\label{FigSM_2D}
\end{figure*}

We now go beyond the adiabatic approximation by considering the full model formed by the coupled Eqs. (1)-(2). As seen in Fig. \ref{Fig3}c, the qualitative shape of the spectrum for the positive mass condensate remains essentially unchanged, although quantitative values of the strength and wavevector for the onset of instability are modified. For the negative mass condensate however, see Fig. \ref{Fig3}d, the spectrum is qualitatively altered compared to the adiabatic approximation, and a novel instability appears. 
Such specifically non-adiabatic instability, not discussed previously in the literature, is \textit{a priori} relevant given our experimental parameters, but is not observed in our measurements.

Several mechanisms can be invoked to explain the suppression of this new instability.
\blue{In particular, in a lattice the polariton linewidth is expected to be minimum in antisymmetric states \cite{Aleiner12,Stepnicki13}. In these states the $\pi$ phase difference of the wavefunction between neighboring sites leads to maximum polariton density in the center of the pillars and zero density in the region in between micropillars. It is in this region that non-radiative losses arising from surface defects generated during the etching process present a higher density. Therefore, antisymmetric states are expected to have the longest lifetime.
In our case (see Fig. \ref{Fig2}b) condensation occurs in such antisymmetric states located in the upper band at the center of the Brillouin zones.
The linewidth is then expected to increase monotonically away from the zone centers as the phase difference between neighboring sites departs from $\pi$, leading to higher probability density in between micropillars and thus broader polariton linewidth.}
In our experiments we can quantify this phenomenon by extracting the linewidth of the low power photoluminescence as a function of the wavevector $k$ (see Supplementary Material, Fig. S5). In the upper band we find an approximate linear increase of the linewidth, $\gamma(k)=\gamma_0+\gamma' \vert k \vert$, with $\hbar\gamma_0=75 \, (130)$ $\mu$eV for high (moderate) $Q$ lattices, and $\gamma'/\gamma_0 \simeq 1.6 \,\mu$m. 
When introducing such momentum-dependent broadening in the simulations, the instability is indeed suppressed, as shown in the spectrum of Fig. \ref{Fig3}f. 
For the 1D wire cavities, photoluminescence measurements also reveal a slight $k$-dependency of the linewidth, $\gamma'/\gamma_0 \simeq 0.7 \,\mu$m, which could be linked to energy relaxation effects \cite{Anton13}.
For completeness, we include this effect in the simulations of the positive mass condensate as well: as shown in Fig. \ref{Fig3}e, the condensate remains in the unstable regime.

To go beyond the stability analysis and simulate the spatial and temporal coherence of the condensate, we now consider the full nonlinear model of Eqs.~(1)-(2), including the $k$-dependent linewidth and the pump profile. We add a gaussian noise term to Eq. (1) so as to effectively account for all quantum, thermal or pump laser fluctuations, as well as a disorder potential with a standard deviation of $30 \, \mu$eV corresponding to the typical disorder strength of our cavities~\cite{Baboux16}. The shape of the disorder is adjusted to fit the experimentally observed condensate density profiles (e.g. Fig. \ref{Fig1}f and j). Figures \ref{Fig1}g and \ref{Fig1}k show the simulated time-averaged condensate density (black) and spatial coherence at $t=0$ (red) for the positive mass condensate, for moderate and high $Q$ wire cavities. Figure \ref{Fig1}m shows the corresponding temporal coherence, all in good agreement with the experiment.

The observed difference between moderate and high $Q$ cavities can be intuitively understood from the interplay between instability and disorder. At moderate $Q$, density fluctuations chaotically propagate along the condensate, resulting in a strongly reduced spatial and temporal coherence: the non-monotonicity in the temporal coherence (Fig. \ref{Fig1}l and m, green line) arises from the scattering of density fluctuations on the disorder. At high $Q$ on the contrary, due to the higher disorder/linewidth ratio, density fluctuations are pinned into localized high density areas, which strongly constraints their dynamics and thus their ability to spoil the temporal coherence (Fig. \ref{Fig1}l and m, blue line). 
\blue{Note that a similar interplay between nonlinearity, dissipation and disorder is encountered in the instability of multimode fibers \cite{Wright16} and random lasers \cite{Mujumdar07}.}

Now turning to negative mass condensates, we calculate their spatio-temporal coherence by considering the highest energy band of the Lieb lattice in the tight-binding limit \cite{Stepnicki13}, and we introduce disorder with the same amplitude than for positive mass. Figures~\ref{Fig2}g and~\ref{Fig2}i show the spatial coherence at zero delay ($g^{(1)}(x,0)$) and the time decay of the coherence ($g^{(1)}(0,t)$). The simulations show a smooth density profile and a high spatial and temporal coherence, in good agreement with the experimental results (Fig.~\ref{Fig2}f,h).

\section{Experimental results in 2D cavities}

The physical mechanism of the polariton instability is expected to be independent of the dimensionality of the system (except at 0D where any dynamics is quenched) \cite{Wouters07}. We have reported above a complete set of data obtained in 1D systems, motivated by the convenience in measuring and calculating coherence properties in these systems, but similar results are expected in 2D cavities.
To test this prediction we have performed spatial coherence measurements in 2D cavities, as reported in Fig. \ref{FigSM_2D}. 
The exciton detuning, polariton mass and normalized pump power $P/P_{\rm th} \approx 2$ are similar to those used for 1D cavities.

The first two columns of Fig. \ref{FigSM_2D} compare results obtained respectively in planar cavities of moderate and high $Q$ factor, excited by a circular Gaussian pump spot with $60$ $\mu$m radius (we used a smaller size spot than for 1D experiments because of power density limitation). From top to bottom are shown: the real space image of the condensate, the interferogram obtained by superimposing two mirror-symmetric images, and profiles of the condensate density and coherence along the axes indicated by blue horizontal lines.

For the moderate $Q$ planar cavity (first column), the condensate extends on about $25 \times 25 $ $\mu$m$^2$ and shows slight density inhomogeneities (black line in Fig. \ref{FigSM_2D}c), with a contrast $\mathcal{C}\simeq 2$.
The spatial coherence (red line in Fig. \ref{FigSM_2D}c) extends on a much shorter scale ($l_c \sim 6 \,\mu$m) than the condensate density, which indicates the presence of an instability with signatures comparable to the 1D moderate $Q$ wire cavity studied in Fig. 1f.

Now turning to measurements performed in a high $Q$ planar cavity (second column of Fig. \ref{FigSM_2D}), we observe a strong increase of the density inhomogeneity of the condensate, with a typical contrast $\mathcal{C}\simeq 10 $. The condensate fragments into distinct lobes, with no mutual coherence between them (Fig. \ref{FigSM_2D}f, red): the coherence length is of the order of the lobe size, $l_c \simeq 10 $ $\mu$m.
These data corroborate the evidence for a fragmentation regime of the polariton instability at high $Q$ factors, similarly to the 1D case (Fig. 1j).

Finally, in the last column of Fig. \ref{FigSM_2D} we investigate condensation in a negative mass state, by using a 2D honeycomb lattice of micropillars (high Q factor and interpillar spacing 2.4 $\mu$m) \cite{Jacqmin14}. Condensation here occurs at the top edge of the $\pi^{\star}$ band, thus in a negative mass quantum state. The honeycomb-shaped spatial distribution of the condensate reflects the lattice geometry. Here we integrate the density and coherence over the three central pillar rows (see horizontal lines).
The envelope slowly decays following a smooth Gaussian-like profile (black line in Fig. \ref{FigSM_2D}i), despite the presence of disorder with a similar amplitude than in the planar cavities.
The spatial coherence (red points in Fig. \ref{FigSM_2D}f) extends over the whole condensate, with a coherence length $l_c \simeq 35$ $\mu$m about $4$ times higher than for positive mass condensates (Fig. \ref{FigSM_2D}c and f), pointing out a stable regime of condensation. Note that the size of the condensates cannot be made larger in these experiments because of limited available excitation density.

\section{Conclusion}

In summary, we have reported a comprehensive study of polariton condensation in III-V cavities, by varying the sign of the effective mass, the dimensionality and the cavity quality factor. 
Due to effective attractive interactions mediated by the exciton reservoir, positive mass condensates are dynamically unstable as evidenced by a strongly reduced spatial and temporal coherence and a spatially inhomogeneous density. Using a lattice to invert the sign of the polariton mass allows to suppress this instability and prepare extended condensates with high spatial and temporal coherence. This method opens exciting possibilities in view of investigating novel driven-dissipative phenomena \cite{Sieberer13,Chiocchetta13,He15,Altman15,Dagvadorj15,Gladilin14,Ji15,Squizzato17} with bosonic condensates in a controlled platform.

\section*{Funding}

This work was supported by the French National Research Agency (ANR) project ”Quantum Fluids of Light” (ANR-16-CE30-0021) and program Labex NanoSaclay via the project ICQOQS (Grant No. ANR-10-LABX-0035), the French RENATECH network, the ERC grant Honeypol and the EU-FET Proactive grant AQUS (Project No. 640800), the Austrian Science Fund (FWF) through SFB FOQUS F40, DK CoQuS W 1210 and the START grant Y 591-N16.

\bigskip

\noindent See Supplementary Material for supporting content.

\bibliography{Biblio}

\begin{thebibliography}{10}
\bibitem{Remark}
More formally, three independent analytical conditions need to be fulfilled so
  as to apply the adiabatic approximation, as shown in Ref. \cite{Bobrovska15}
  .
\bibitem{Remark2}
\blue{In our lattice we note that due to the large interpillar spacing
  (required to obtain a band with large negative mass) the optical overlap
  between pillars is small. In this limit, the additional Lieb sites are only
  weakly coupled and play a minor role, so that the band structure resembles
  that of a simple linear chain of pillars of period $a'=a/2$. In the case of
  such a simple linear chain there is a single band that takes a negative
  curvature at the edge of the Brillouin zone, located at $k=\pm \pi/a'=\pm
  2\pi/a$, i.e. at the same absolute wavevector position than for the Lieb
  lattice we consider here} .
\bibitem{Remark3}
\blue{The density of the interference fringes is fixed by the relative angle
  between both arms of the interferometer. In Figs. 3d and 5h we used for
  convenience a higher fringe density to be able to measure a sufficient number
  of fringes within the small spatial areas occupied by the condensate. This
  does not affect the fringe contrast, which is the only relevant quantity to
  extract the coherence} .
\end{thebibliography}

\end{document}